\documentclass[aps,showpacs,nobibnotes, twocolumn,
nobalancelastpage]{revtex4}
\usepackage{graphicx}
\usepackage{graphics}
\usepackage{amsmath}
\usepackage{amssymb}
\usepackage{latexsym}
\usepackage{dcolumn}% Align table columns on decimal point
\usepackage{bm}% bold math
\newcommand{\simgt}{\lower.5ex\hbox{$\; \buildrel > \over \sim \;$}}
\newcommand{\simlt}{\lower.5ex\hbox{$\; \buildrel < \over \sim \;$}}
\newcommand{\be}{\begin{equation}}
\newcommand{\ee}{\end{equation}}
\newcommand{\ba}{\begin{eqnarray}}

\newcommand{\bi}{\begin{itemize}}
\newcommand{\ei}{\end{itemize}}

\newcommand{\bfi}{\begin{figure}
\epsfxsize=9cm 
\epsffile}
\newcommand{\efi}{\end{figure}}

\newcommand{\mnras}{MNRAS}

\newcommand{\apjs}{ApJS}

\DeclareFontFamily{OT1}{rsfs10}{} 
\DeclareFontShape{OT1}{rsfs10}{m}{n}{ <-> rsfs10 }{} 
\DeclareMathAlphabet{\mathscript}{OT1}{rsfs10}{m}{n} 

\def\d{\delta}

\def\k{\kappa}
\def\l{\lambda}

\def\p{\pi}

\def\r{\rho}

\def\L{\Lambda}
\def\O{\Omega}

\begin{document}
\title{Three-Point Correlations in f(R) Models of Gravity}
\author{Alexander Borisov \& Bhuvnesh Jain}
\email{borisov@physics.upenn.edu}
\affiliation{
Department of Physics and Astronomy, University of Pennsylvania,
  Philadelphia, PA 19104\\
}

\begin{abstract}
Modifications of general relativity provide an alternative explanation
to dark energy for the observed acceleration of the universe.  
We calculate quasilinear effects in the growth of structure in $f(R)$
models of gravity using perturbation theory. 
We find significant deviations in the bispectrum that depend on cosmic time,
length scale and triangle shape. However the deviations in the 
reduced bispectrum $Q$ for $f(R)$ models are at the
percent level, much smaller than the deviations in the bispectrum
itself.  This implies that three-point correlations 
can be predicted to a good approximation
simply by using the modified linear growth factor in the standard
gravity formalism. Our results suggest  that gravitational clustering in the
weakly nonlinear regime is not fundamentally altered, at least for a class of
gravity theories that are well described in the Newtonian regime
by the parameters $G_{\rm eff}$ and $\Phi/\Psi$. This approximate
universality was also seen in   
the N-body simulation measurements of the power spectrum by Stabenau \&
Jain (2006), and in other recent studies based on simulations. 
Thus predictions for such modified gravity models in
the regime relevant to large-scale structure observations
may be less daunting than expected on first principles. 
We discuss the many caveats that apply to such predictions. 
\end{abstract}
\maketitle

\section{Introduction}

The energy contents of the universe pose an interesting puzzle, 
in that general relativity (GR) plus the Standard Model of particle
physics can only account for about $4\%$ of the energy density inferred from
observations.  By introducing dark matter and dark  
energy, which account for the remaining $96\%$ of the total 
energy budget of the universe,  cosmologists have been able to
account for a wide range of  observations, from  
the overall expansion of the universe to 
the large scale structure of the early and late
universe~\cite{Reviews}. 

The dark matter/dark energy scenario assumes the validity of GR at
galactic and cosmological scales and introduces exotic components of
matter and energy to account for observations. Since 
GR has not been tested independently on these scales, a natural
alternative is that GR itself needs to be modified on large scales. 
This possibility, that modifications in
GR on galactic and cosmological
scales can replace dark matter and/or dark energy,  
has become an area of active research in recent years. 

Attempts have been made to modify GR with a focus on 
galactic~\cite{MOND} or cosmological
scales~\cite{DGP,fR,Sahni2005}.   Modified Newtonian Dynamics (MOND) and its
relativistic version (Tensor-Vector-Scalar, TeVeS) ~\cite{MOND} 
attempt to explain observed galaxy rotation curves without dark matter
(but have problems on larger scales). The DGP model~\cite{DGP}, in which
gravity lives in a 5D brane world,  naturally leads to late time 
acceleration of the universe. 

Adding a correction term $f(R)$ to the
Einstein-Hilbert action \cite{fR} also allows late time acceleration of
the universe   to be realized.  

In this paper we will focus on modified gravity (MG) theories that are
designed as an alternative to dark energy to produce the present day 
acceleration of the universe. In these models, such as DGP and $f(R)$
models, gravity at late cosmic times and on large-scales departs from
the predictions of GR.  By design, successful MG models
are difficult to distinguishable from viable DE models against observations of
the expansion history  of the universe. However, in general they
predict a different growth of perturbations which can be tested using
observations of large-scale structure (LSS) 
\cite{Yukawa,Stabenau2006,Skordis06,Dodelson06,DGPLSS,consistencycheck,Koyama06,fRLSS,Zhang06,Bean06,MMG,Uzan06,Caldwell07,Amendola07, HuSaw3}. 

In this paper we consider the quasilinear regime of clustering in
which perturbation theory calculations are valid. We 
explore what $f(R)$ modifications to
gravity predict about the behavior of the three-point
correlation function. In \S II we outline the
particular type of $f(R)$ gravity model we will be using for our
calculations. In \S III we introduce the fundamentals of
perturbation theory and in particular how it applies to modified
gravity. In \S IV we focus on second order corrections and in
particular the Bispectrum. In \S V we present our results and compare
with other studies of the nonlinear regime. 
In \S VI we discuss the implications for observations.

\section{The $f(R)$ modified gravity model}

In general $f(R)$ models are a modification of the Einstein-Hilbert action of the form:

\begin{equation}
 S = \int d^{4}x \sqrt{-g} \left[ \frac{R + f(R) }{2 \k^{2} } + {\cal L}_{m} \right],
\end{equation}
where $R$ is the curvature, $\k^{2} = 8 \p G$, and ${\cal L}_{m}$ is
the matter Lagrangian. A major issue with gravity modifications has
been that while they are successful at explaining the acceleration of
the universe, they also tend to fail to comply with Solar system (very
small scales) observations. Recently, though, the so called Chameleon
mechanism was found \cite{Khoury1} that makes it possible to overcome
this problem. Significant attempts have been made to include Chameleon
behavior \cite{KhouryDGP} in DGP theories as well. 
One example of an $f(R)$ model that exhibits Chameleon behavior was
constructed by Hu and Sawicki \cite{HuSaw}. The functional form of
$f(R)$ there is derived from a list of observational requirements: it
should mimic $\L CDM $ in the high-redshift regime as well as produce $\L CDM $
acceleration of the universe at low redshift without a true
cosmological constant; it should also fit Solar system observations. 

The particular form chosen by \cite{HuSaw} is:
\begin{equation}
 f(R) = - m^{2} \frac{c_{1} (R/m^{2})^{n}}{c_{2} (R/m^{2})^{n} + 1}
\end{equation}
with
\begin{equation}
 m^{2} \equiv \frac{\k^{2}\bar{\r}_{0}}{3} = (8315 Mpc)^{-2} \left( \frac{\O_{m} h^{2}}{0.13} \right)
\end{equation}
where $\kappa \equiv 8\pi G$ and  $\bar{\r}_{0}$ is the average density today.
In this model, modifications to GR only appear at low redshift, when 
 we are safely in the matter dominated regime. The properties of the
 model are well described by the auxiliary scalar field $f_{R} \equiv
 \frac{df(R)}{dR}$. 

 Before going to the expansion history, it is worth briefly reviewing
 the main features of this model, following the 
original presentation in \cite{HuSaw}. The trace of the modified
 Einstein equations serves as the equation of motion for $f_{R}$: 
\begin{equation}
\label{fielde}
 3 \Box f_{R} - R + f_{R}R -2f = -\k^{2} \r , 
\end{equation}
or, in  terms of the effective potential, 
\begin{equation}
 \Box f_{R} = \frac{\partial V_{\rm eff}}{\partial f_{R}} . 
\end{equation}
The effective mass for the $f_{R}$ field is then given by the second
derivative of $V_{\rm eff}$, evaluated at its extremum:
\begin{equation}
 m^{2}_{\rm eff} = \frac{\partial^2 V_{\rm eff}}{\partial f^2_{R}} =
 \frac{1}{3} \left( \frac{1+f_R}{f_{RR}} -R \right) . 
\end{equation}
The Compton wavelength of the field is then given by 
%\begin{equation}
$ \l_{f_R} \equiv m^{-1}_{\rm eff} $. 
%\end{equation}

It is very convenient to introduce a dimensionless quantity:
\begin{equation}
 B = \frac{f_{RR}}{1+ f_{R}} R^{'} \frac{H}{H^{'}}
\end{equation}
It has been shown \cite{HuSaw} that in the high-curvature regime $B$ 
is connected to the Compton wavelength via:
\begin{equation}
 B^{1/2} \sim  \l_{f_R} H
\end{equation}
and thus is essentially the Compton wavelength of $f_R$ at the
background curvature in units of the horizon length. 

In the static limit with $|f_R| \ll 1$ and $|f/R| \ll 1$,
Eqn. \ref{fielde} becomes 
\begin{equation}
 \bigtriangledown^2 f_R \approx \frac{1}{3} (R - \k^2 \r )
\end{equation}
where $\r$ is the local density. This equation has 2 modes of
solutions. One is the very high curvature $R \approx \k^2  \r$ and the
other one is the low curvature (but still high compared to the
background density) $R \ll \k^2 \r$. For more on the interplay of
these two regimes and applications in solar system observations see
\cite{HuSaw}.  

Let's now move on to the expansion history. For the model to yield
behavior that is observationally viable requires a choice for the
present day value of the $f_R$ field $f_{R0} \ll 1$. 
This is equivalent to $R_0 \gg m^2$. In that case the
approximation $R \gg m^2$ is valid for the whole expansion history and
we have: 

\begin{equation}
 \lim_{m^2/R \rightarrow 0} f(R) \approx -\frac{c_1}{c_2}m^2 + \frac{c_1}{c^2_2}m^2 \left( \frac{m^2}{R} \right)^n
\end{equation}

In the limiting case of $ c_{1} / c^{2}_{2} \rightarrow 0 $ at fixed $c_1 / c_2$ we obtain a cosmological constant behavior $\L CDM$. Thus to approximate the $\L CDM$ expansion history with a cosmological constant $\O_{\L}$ and matter density $\O_m$ we set:

\begin{equation}
\frac{c_1}{c_2} \approx 6 \frac{\O_{\L}}{\O_m}
\end{equation}

This leaves 2 remaining parameters to play with: $n$  and $ c_{1} / c^{2}_{2}$ to control how closely the model mimics $\L CDM$. Larger n mimics until later in the expansion history, while smaller $ c_{1} / c^{2}_{2}$ mimics it more closely.

For flat $\L CDM$ we have the following relations:

\begin{equation}
 R \approx 3m^2 \left( a^{-3} +4 \frac{\O_{\L}}{\O_m} \right)
\end{equation}

\begin{equation}
 f_R = -n \frac{c_1}{c^2_2} \left( \frac{m^2}{R} \right)^{n+1}
\end{equation}

As we will see later these are the necessary ingredients for the
application of perturbation theory to the model. 

Finally we need to obtain a suitable parametrization of the mode. At the present epoch we have:
\begin{equation}
 R_0 \approx m^2 \left(\frac{12}{\O_m} -9 \right)
\end{equation}

\begin{equation}
 f_{R0} \approx -n \frac{c_1}{c^2_2} \left(\frac{12}{\O_m} -9 \right)^{-n-1}
\end{equation}

In particular, for $\O_{\L}=0.76$ and $\O_m = 0.24$, we have $R_0=41m^2$
and $f_{R0} \approx -n\ c_1 / c^2_2 / (41)^{n+1} $. 
>From now on we will parametrize the model through
$f_{R0}$ and $n$. Fig. 9 in \cite{HuSaw} shows that there is a wide
range of viable parameter values  which satisfy Solar system and
Galaxy requirements. We re-iterate that what makes this
particular $f(R)$ model viable is
its Chameleon behavior -- the possibility of uniting galaxy and
solar system observations with the expansion of the Universe. 

\section{Perturbation Formalism}
\label{sec:formalism}

By definition, the dark sector (dark matter and dark energy) can only
be inferred from  its gravitational consequences.
In general relativity,  gravity is determined by the total
stress-energy tensor of all matter and energy ($G_{\mu\nu}=8\pi
G\ T_{\mu\nu}$).  

We may consider the Hubble parameter $H(z)$ to be fixed by observations. 
In a dark energy model, $\bar{\rho}$ is given by the Friedman
equation of GR:  $\bar{\rho}=3H^2/8\pi G$.  The equation of
state parameter is $w=-1-2\dot{H}/3H^2$. 
The corresponding modified gravity model has matter
density to be determined from its Friedman-like
equation. We will consider MG models dominated by dark matter and
baryons at late times. 

\subsection{Metric and fluid perturbations}

With the smooth variables fixed, we will consider perturbations as
a way of testing the models. 
In the Newtonian gauge, scalar perturbations to the metric
are fully specified by two scalar potentials $\Psi$ and $\Phi$:
\begin{equation}
ds^2 = -(1-2\Psi)\ dt^2 + (1-2\Phi)\ a^2(t)\ d{\vec x}^2 
\label{eqn:metric}
\end{equation}
where $a(t)$ is the expansion scale factor. This form for the
perturbed metric is fully general for any metric theory of
gravity, aside from having excluded vector and tensor perturbations
(see \cite{Bertschinger2006} and references therein for justifications). 
Note that $\Psi$ corresponds to the Newtonian potential for
the acceleration of particles, and that in general relativity
$\Phi=-\Psi$ in the absence of anisotropic stresses.  

A metric theory of gravity relates the two potentials above to the
perturbed energy-momentum tensor. We introduce variables to
characterize the density and velocity perturbations for a fluid, which
we will use to 
describe matter and dark energy. 
The density fluctuation $\delta$ is given by
\begin{equation}
\delta({\vec x},t) \equiv \frac{\rho({\vec x},t) -
  {\bar\rho(t)} } {\bar\rho(t)} 
\label{eqn:delta}
\end{equation}
where $\rho({\vec x},t)$ is the density and ${\bar\rho(t)}$ is the cosmic
mean density. The second fluid variable is the divergence of the
peculiar velocity 
\begin{equation}
\theta\equiv\nabla_j T_0^j/(\bar{p}+\bar{\rho})={\vec \nabla} \cdot {\vec v}, 
\end{equation}
where $\vec v$ is the (proper) peculiar velocity.  
Choosing
$\theta$ instead of the vector ${\bf v}$ implies that we have assumed
${\bf v}$ to be irrotational. Our notation and formalism follows that
of \cite{JZ}. 
%This approximation is sufficiently
%accurate in the linear regime, even for unconventional dark energy
%models and minimally coupled modified gravity models.

In principle, observations of large-scale structure can directly 
measure the four perturbed variables introduced above: 
the two scalar potentials $\Psi$ and $\Phi$, and the density and velocity
perturbations specified by $\delta$ and $\theta$. 
It is convenient to work with the Fourier transforms, such as: 
\begin{equation}
\hat\delta(\vec k,t) = \int d^3 x \ \delta(\vec x,t) \
e^{-i {\vec k} \cdot{\vec x}} 
\label{eqn:FT}
\end{equation}
When we refer to length scale $\lambda$, it corresponds to a
a statistic such as the power spectrum on wavenumber $k=2\pi/\lambda$. 
We will henceforth work exclusively with the Fourier space quantities 
and drop the $\hat{}$ symbol for convenience. 

\subsection{Linearized fluid equations}
We will use the perturbation theory equations for
the quasi-static regime of the growth of perturbations. 
%Through the interpolation function of Hu
%and Sawicki, we will extend the solutions towards the super-horizon up to
%scales at which the assumptions under which the quasi-static
%perturbation theory equations are derived are valid.  
We begin with the fluid equations in the  Newtonian gauge,  
%($ds^2=(1+2\Psi)dt^2-a^2(1-2\Phi)d{\bf  x}^2$) 
following the formalism and notation of \cite{Ma95}. 

For minimally coupled gravity models with baryons and cold dark
matter, but without dark energy, we can neglect pressure and anisotropic
stress terms in the evolution equations to get the continuity equation: 
\be
\label{eqn:MG1}
%{\bf MG: }
\dot{\delta}=-\left(\frac{\theta}{a}-3\dot{\Phi}\right)\simeq -
\frac{\theta}{a} \ , 
\ee
where the second equality follows from the quasi-static
    approximation as for GR. The Euler equation is: 
\be
\label{eqn:MG2}
%{\bf MG: }
\dot{\theta}=-H\theta - \frac{k^2\Psi}{a} \ .
\ee

%We will attempt to characterize the general behavior in the weak field
%limit for small perturbations (small $\delta$) 
%and non-relativistic motions. 
We parametrize modifications in gravity by two functions 
$\tilde{G}_{\rm  eff}(k,t)$ and $\eta(k,t)$ to get the analog of the
Poisson equation and a second equation connecting $\Phi$ and
$\Psi$~\cite{Zhang7,JZ}. We first write the generalization of the
Poisson equation in terms of an effective gravitational constant
$G_{\rm eff}$:
\be
\label{eqn:MG3a}
k^2 \Phi=-4\pi G_{\rm eff}(k,t)\bar{\rho}_{\rm MG}a^2\delta_{\rm
  MG}\ .
\ee
Note that the potential $\Phi$ in the Poisson equation comes from the
  spatial part of the metric, whereas it is the ``Newtonian'' potential $\Psi$
  that appears in the Euler equation (it is called the
  Newtonian potential   as its gradient gives the acceleration of
  material particles). Thus in MG, one cannot directly use the Poisson
  equation to eliminate the potential in the Euler equation. 
  A more useful version of the Poisson 
  equation would relate the sum of the potentials,
  which determine lensing, with the mass density. We therefore
  introduce $\tilde G_{\rm eff}$ and write the constraint equations for
  MG as
\be
\label{eqn:MG3}
k^2(\Phi-\Psi)=-8\pi \tilde{G}_{\rm eff}(k,t)\bar{\rho}_{\rm MG}a^2\delta_{\rm
  MG}
\ee
\be
\label{eqn:MG4}
\Phi=-\Psi \ \eta(k,t)
\ee
where $\tilde{G}_{\rm eff} = G_{\rm eff}(1+\eta^{-1})/2$.

\begin{figure}[tbp]
\renewcommand{\baselinestretch}{1}
\begin{center}
\leavevmode

\begin{minipage}[b]{.45\textwidth}
\includegraphics[width=1\textwidth]{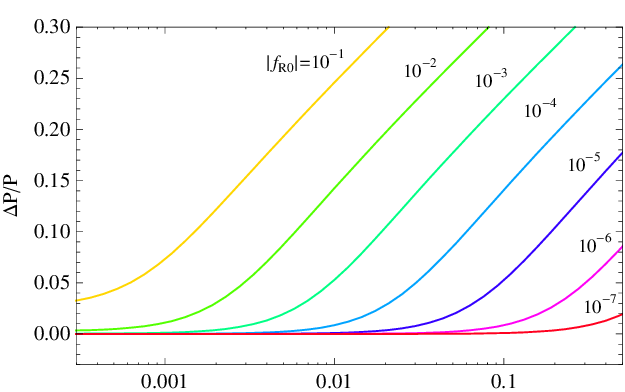}
\end{minipage}%

\vspace{-10pt}
\begin{minipage}[b]{0.45\textwidth}
\includegraphics[width=1\textwidth]{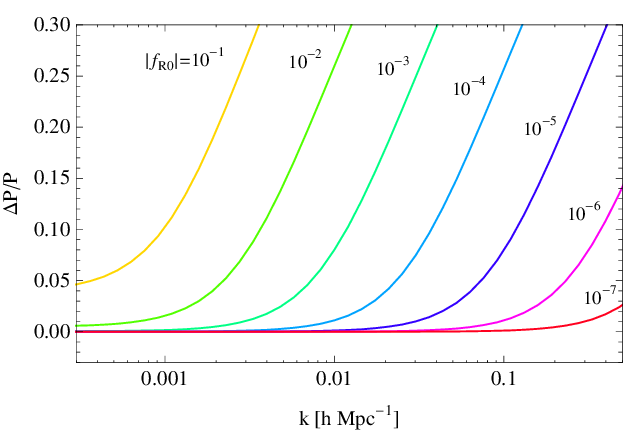}
\end{minipage}

\end{center}

\caption{\label{fig:fig1}{\footnotesize Fractional
    change  at $z=0$ in   
    the density power spectrum of the $f(R)$ model compared to $\L
    CDM$ for a set of choices of $f_{R0}$. The upper panel is the
    $n=4$ model, while the lower panel has $n=1$. This figure can be
    compared with Fig. 4 in    \cite{HuSaw}}. }
\end{figure}

\begin{figure}[tbp]
\renewcommand{\baselinestretch}{1}
\begin{center}
\leavevmode

\begin{minipage}[b]{.45\textwidth}
\includegraphics[width=1\textwidth]{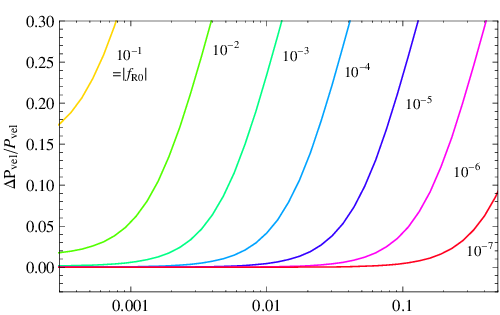}
\end{minipage}%

\vspace{-10pt}
\begin{minipage}[b]{.45\textwidth}
\includegraphics[width=1\textwidth]{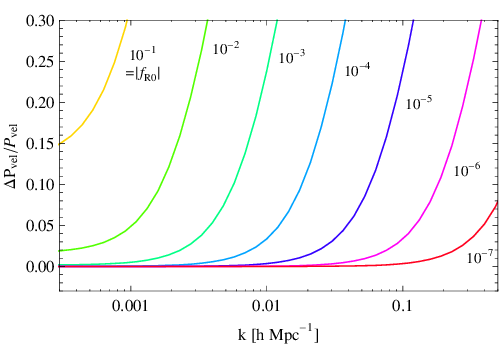}
\end{minipage}%

\end{center}

\caption{\label{fig:fig1A}{\footnotesize Fractional
    change  at $z=0$ in the velocity power spectrum of the $f(R)$
    model compared to $\L CDM$ for a set of choices of
    $f_{R0}$. The upper panel is the
    $n=4$ model, while the lower panel has $n=1$. }}
\end{figure}

The parameter $\tilde{G}_{\rm eff}$ characterizes
deviations in the ($\Phi-\Psi$)-$\delta$ relation from that in GR. Since the
combination $\Phi-\Psi$ is directly responsible for gravitational
lensing, 
$\tilde{G}_{\rm eff}$ has a specific physical meaning: it determines the
power of matter 
inhomogeneities to distort light. This is the reason we prefer it over
working with more direct generalization of Newton's constant,
${G}_{\rm eff}$. 

With the linearized equations above, the evolution of either the
density or velocity perturbations can be described by a single second
order differential equation. 
%In the case of MG theories, this equation
%is simpler as the only source is provided by the Newtonian potential
%$\Psi$. 
 From  Eqns. \ref{eqn:MG1} and \ref{eqn:MG2} we get, for the linear
solution, $\delta(\vec k,t)\simeq \delta_{initial}(\vec k) D(k,t)$, 
\begin{equation}
\ddot{\delta}+2 H \dot{\delta} + \frac{k^2 \Psi}{a}
 = 0 . 
\label{eqn:lingrowth}
\end{equation}
For a given theory, Eqns. \ref{eqn:MG3} and \ref{eqn:MG4} then
allow us to substitute for $\Psi$ in terms of $\delta$ to
determine $D(k,t)$, the linear growth factor for the density: 
\begin{equation}
\ddot{D}+2 H \dot{D} - \frac{8 \pi \tilde{G}_{\rm eff}}{(1+\eta)}
\bar{\rho}a^2\ D = 0 . 
\label{eqn:lingrowth2}
\end{equation}

What we need now are $\tilde{G}_{\rm eff}$ and $\eta$ for our modified gravity model.

\subsection{Parametrized post-Friedmann framework}
%The parametrized post-Friedmann framework}

Hu \& Sawicki \cite{HuSaw} also developed a formalism for simultaneous
treatment of the super-horizon and quasi-static regime for modified
gravity theories (in particular $f(R)$ and DGP) 
%called parameterized
%post-Friedmann framework (PPF) for modified gravity
\cite{HuSaw2}. They begin by describing the different regimes
individually and the requirements they impose on the structure of such
models. Consequently they describe a linear theory parametrization of
the super-horizon and the quasi-static regime and test it against
explicit calculation. What is important for this paper is the proposed
interpolation function for the metric ratio  

\begin{equation}
g= \frac{\Phi+\Psi}{\Phi-\Psi}
\end{equation}

For consistency we will express formulae from Hu \& Sawicki 
in terms of physical time instead of expansion factor. 
We start with a background FRW universe for which we have the
curvature in terms of the Hubble parameter $R = 6\dot{H}+ 12H^{2}$. As
we mentioned the background evolution $H$ is chosen to match that of  
flat $\L CDM$. We can then compute the Compton parameter $B$ for our preferred
model since we know what $f(R)$, $R$ and $H$ are: 

\begin{equation}
 B = \frac{f_{RR}}{1+ f_{R}} \dot{R} \frac{H}{\dot{H}}
\end{equation} 

The next step is to look at the super-horizon regime. In that case we
know how to calculate the potentials $\Phi$ and $\Psi$. Their
evolution is given by \cite{HuSaw3}: 
\begin{equation}
\begin{gathered}
 \ddot{\Phi} + \left( 1 - \frac{\ddot{H}}{H \dot{H}} + \frac{\dot{B}}{H(1-B)} +B\frac{\dot{H}}{H^2} \right) H \dot{\Phi} + \\
 \left( 2\frac{\dot{H}}{H^2} - \frac{\ddot{H}}{H \dot{H}} + \frac{\dot{B}}{H(1-B)} \right) H^2 \Phi = 0, (k_{H} \rightarrow 0)
\end{gathered}
\end{equation}
\begin{equation}
 \Psi = \frac{-\Phi - B \dot{\Phi}/H}{1-B} , (k_{H} \rightarrow 0)
\end{equation}
This allows us to compute %the following quantity: 
$ g_{SH} = g(t, k_{H}\rightarrow 0)$. 
%\begin{equation}
% g_{SH} = g(t, k_{H}=0) = \frac{\Phi+\Psi}{\Phi-\Psi} , (k_{H} \rightarrow 0),
%\end{equation}

Furthermore  in the case of subhorizon evolution where
$k_{H}=k/a(t)H \gg 1$ we have $g_{QS}=-\frac{1}{3}$ \cite{HuSaw3}. 

According to \cite{HuSaw2} in the case of $f(R)$ theories we can use the following interpolation functions for the metric ratio:
\begin{equation}
g(t, k)= \frac{g_{SH} + g_{QS}(c_{g}k_{H})^{n_{g}}}{1+(c_{g}k_{H})^{n_{g}}}
\end{equation}
The evolution then is well described by \cite{HuSaw2} $c_{g}=0.71B^{1/2}$ and $n_{g}=2$
In terms of the post-Newtonian parameter: $\eta = -\Phi/\Psi$ we have $g = (\eta-1)/(\eta+1)$.

We still need one more ingredient, $\tilde{G}_{\rm 
eff}$, which is given by \cite{HuSaw2} 
\begin{equation}
\label{Gteff}
\tilde{G}_{\rm eff} (k,t)= \frac{G}{1+f_{R}} 
\end{equation}

We now have all the needed components to use in equation
(\ref{eqn:lingrowth2}) for the growth factor of density perturbations.
Calculations of  the fractional change in the linear density power spectrum
compared to GR have been done \cite{HuSaw}. We show these in 
Fig. \ref{fig:fig1}, using $\tilde{G}_{\rm eff}$ and $\eta$ 
for the particular model we
investigate. We have assumed the transfer function for the concordance
 $\Lambda-$CDM model consistent with the 5-year WMAP data \cite{2008arXiv0803.0586D}.

We can also use the relations given above to obtain the linear growth
factors for $\theta$ and the potentials from $D$. Note that in general 
the growth factors for the potentials have a different $k$ dependence
than $D$. 

\section{The Bispectrum in Perturbation Theory}

\subsection{Nonlinear Fluid Equations}

The fluid equations in the Newtonian regime are given by the
continuity, Euler and Poisson equations. Keeping the nonlinear terms
that have been discarded in the study of linear perturbations above, the continuity equations gives: 
\begin{equation}
\dot{\delta} + \theta = -\int \frac{d^3 k_1}{(2\pi)^3}  \frac{{\vec 
k}\cdot{\vec k_1}}{k_1^2} \theta(\vec{k_1})\delta({\vec k}-{\vec k_1})
\label{eqn:continuityNL}
\end{equation}
where the term on the right shows the nonlinear coupling of
modes. Note that the time derivatives are with respect to conformal
time in this section. 

The Euler equation is
\begin{equation}
\dot{\theta} + H \theta + k^2 \Psi = 
-\int \frac{d^3 k_1}{(2\pi)^3}  
\frac{k^2{\vec k_1}\cdot(\vec k- \vec k_1)}{2 k_1^2 |{\vec k_1} - 
{\vec k_2}|^2 }
\theta(\vec{k_1})\theta({\vec k}-{\vec k_1})
\label{eqn:eulerNL}
\end{equation}
We neglect pressure and anisotropic stress as the energy density is
taken to be dominated by non-relativistic matter \cite{Jain1994}. 
The Poisson equation is given by Eqn. \ref{eqn:MG3} and
supplemented by the relation between $\Psi$ and $\Phi$ given by
Eqn. \ref{eqn:MG4}. Using these equations we can substitute for
$\Psi$ in the Euler equation to get 
\begin{eqnarray}
\dot{\theta} &+&  H \theta + \frac{8 \pi \tilde{G}_{\rm eff}}{(1+\eta)} 
\bar{\rho}_{\rm MG}a^2\delta \nonumber \\
&=& \, -\int \frac{d^3 k_1}{(2\pi)^3}  
\frac{k^2{\vec k_1}\cdot({\vec k- \vec k_1})}{2 k_1^2 |{\vec k_1} - 
{\vec k_2}|^2 }
\theta(\vec{k_1})\theta({\vec k}- {\vec k_1})
\label{eqn:eulerNL2}
\end{eqnarray}
Eqns. \ref{eqn:continuityNL} and \ref{eqn:eulerNL2} are 
two equations for the two variables $\delta$ and
$\theta$. They constitute a fully nonlinear
description and can be solved once $\eta$ and $\tilde{G}_{\rm
  eff}$ are specified. An important caveat is that they may nevertheless be 
invalid on strongly nonlinear scales or for particular MG  theories. 
For the $f(R)$ model considered here, they are valid on scales well above 1
Mpc; on smaller scales the chameleon mechanism modifies the growth of
structure \cite{LimaHu}. Since we will use perturbation theory, our
approach breaks down once $\delta\sim1$ in any case. 

Next we consider perturbative expansions for the density field and the
resulting behavior of the power spectrum and bispectrum. Let $\delta =
\delta_1 + \delta_2 +...$ where 
%$\delta_1 \equiv \delta_1(\vec{k},a=1) \delta_1(\vec{k},t)$ and 
$\delta_2 \sim O(\delta_1^2)$. 
In the quasilinear regime, 
i.e. on length scales between $\sim$10-100 Mpc, mode coupling effects can be
calculated using perturbation theory. 
While this is strictly true only for
general relativity, a MG theory that is close enough to GR to fit
observations can also be expected to have this feature. 

For MG, following  \cite{JZ}, let us simplify the notation by introducing the function: 
\begin{equation}
\zeta(k,t) = \frac{8 \pi \tilde{G}_{\rm eff}}{(1+\eta)} ,  
\label{eqn:MGfunction}
\end{equation}
which is simply $4\pi G$ in GR but can vary with time and scale in MG
theories.  
%Note that for $f(R)$ gravity, we have 
%\begin{equation}
%\zeta(k,t) = 
%8 \pi \frac{1}{1+f_R}\left( \frac{1+2k^2 \xi}{2+3k^2\xi} \right)
%\end{equation}
%where $\xi = 2 f_{RR}/[a^2 (1+f_R)]$ and subscript $_R$ denotes
%derivates with respect to $R$. 
The evolution of the linear growth factor is given by substituting for
$\Psi$ in Eqn. \ref{eqn:growth} to get (as above, but with conformal
time here) 
\begin{equation}
\ddot{\delta_1}+ H \dot{\delta_1} - \zeta \bar{\rho}a^2
\delta_1 = 0 . 
\label{eqn:growth}
\end{equation}
The linear growth factor has a scale dependence for our $f(R)$ as
shown in Fig. 1. 
%In GR, the relation of $\Psi$ to $\delta$ is given by the Poisson
%equation with constant $G$. In MG, this relation involves both $\eta$
%and $\tilde{G}_{\rm eff}$. If either of these functions have a dependence on
%$k$ or $z$, then the solution for the growth factor changes. 
%The linear solutions for $\Psi$ and $\Phi-\Psi$ are then simply
%obtained using  Eqns. \ref{eqn:MG3} and \ref{eqn:MG4}. 

In addition we show below that the second order solution has a
functional dependence on  
$\tilde{G}_{\rm eff}$ and $\eta$ that can differ from GR. 
Thus potentially distinct signatures of the scale and time dependence of 
$\tilde{G}_{\rm eff}(k,z)$ can be inferred from higher order terms. 
These rely either on features in $k$ and $t$ in 
measurements of $P_{\Phi-\Psi}$ and $P_\delta$, or
on the three-point functions which 
can have distinct signatures of MG even at a single
redshift \cite{Bernardeau2004}.  Quasilinear signatures due to
$\eta(k,z)$ can also  
be detected via second order terms in the redshift distortion
relations for the power spectrum and bispectrum. Our discussion
generalizes that of \cite{Yukawa}
who examined a Yukawa-like modification of the Newtonian potential. 
%%BJ: this needs to be checked. 

\subsection{Second order solution}
 From a perturbative treatment of Eqns. \ref{eqn:continuityNL} 
and \ref{eqn:eulerNL2} the second order term for the
growth of the density field is given by \cite{JZ}
\begin{eqnarray}
&&\ddot{\delta_2} + H\dot{\delta_2} - \bar{\rho}a^2
\zeta \delta_2 = \nonumber \\
&& H I_1[\dot{\delta_1}, \delta_1] 
+ I_2[\dot{\delta_1},\dot{\delta_1}] + \dot{I_1}[\dot{\delta_1},  \delta_1] , 
\label{eqn:D2}
\end{eqnarray}
where $I_1$ and $I_2$ denote convolution like integrals of the two
arguments shown, given by the right-hand side of equations 
\ref{eqn:continuityNL} and \ref{eqn:eulerNL2} as follows
\begin{equation}
I_1[\dot{\delta_1},\delta_1](\vec{k}) = \int \frac{d^3 k_1}{(2\pi)^3}
  \frac{{\vec k}\cdot{\vec
    k_1}}{k_1^2} \dot{\delta_1}(\vec{k_1})\delta_1({\vec k}-{\vec k_1})
\end{equation}
and
\begin{equation}
I_2[\dot{\delta_1},\dot{\delta_1}](\vec{k}) = 
\int \frac{d^3 k_1}{(2\pi)^3}  
\frac{k^2{\vec k_1}\cdot(\vec k- \vec k_1)}{2 k_1^2 |{\vec k_1} - 
{\vec k_2}|^2 }
\dot{\delta_1}(\vec{k_1})\dot{\delta_1}({\vec k}-{\vec k_1}) .
\end{equation}
Finally, the last term in Eqn. \ref{eqn:D2} is simply 
$\dot{I_1}[\dot{\delta_1},  \delta_1] = I_1[\ddot{\delta_1}, \delta_1] +
I_1[\dot{\delta_1},\dot{\delta_1}]$. Note that by continuing the iteration, 
higher order solutions can be obtained.

\begin{figure}[tbp]
\renewcommand{\baselinestretch}{1}
\begin{center}
\leavevmode
\begin{minipage}[b]{.45\textwidth}
\includegraphics[width=1\textwidth]{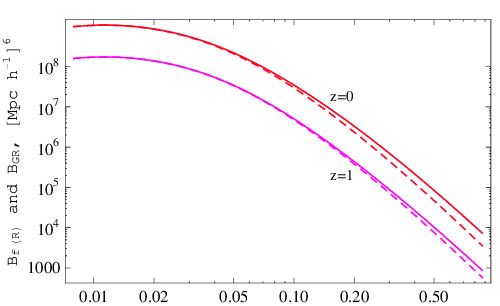}
\end{minipage}%

\vspace{-10pt}
\hspace{2pt}
\begin{minipage}[b]{.422\textwidth}
\includegraphics[width=1\textwidth]{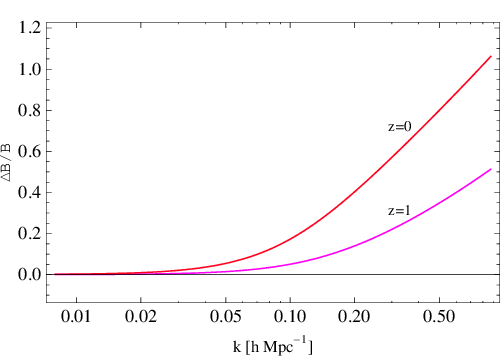}
\end{minipage}
\end{center}

\caption{\label{fig:fig3}{\footnotesize {\it Upper panel:} The
    Bispectrum of the $f(R)$ model for equilateral triangles depending
    on scale for the $f(R)$ model with $f_{R0}=10^{-5}$, $n=1$. The
    corresponding regular gravity bispectrum is shown as the dashed
    curve. {\it Lower panel:} The Ratio of the f(R) Bispectrum to the
    regular gravity one.}} 
\end{figure}

\subsection{Three-point correlations}

Distinct quasilinear effects are found in three-point correlations -- 
we will use the Fourier space bispectrum. Recently Tatekawa \& 
Tsujikawa have performed a similar perturbative analysis and presented
results on the skewness \cite{TatekawaTsujikawa}. We prefer to use the
bispectrum as it allows one to study specific quasilinear signatures
contained in the configuration dependence of triangle shapes. The
bispectrum for the density field $B_\delta$ is defined by
\begin{equation}
\langle \delta({\vec k_1}) \delta({\vec k_2})\delta({\vec k_3}) \rangle = 
(2 \pi)^3 \delta_{\rm D}({\vec k_1 + \vec k_2 + \vec k_3}) 
B_\delta(\vec k_1, \vec k_2, \vec k_3)
\label{eqn:bispectrum}
\end{equation}
Since $B_\delta\sim \langle \delta^3\rangle\sim \langle \delta_1^2
\delta_2\rangle$ (using $\langle\delta_1^3\rangle=0$ for an initially
Gaussian density field) at tree level, 
the second order solution enters at leading order in the
bispectrum. Note also that the wavevector arguments of the bispectrum
form a triangle due to the Dirac delta function on the right-hand
side above.

The bispectrum is the lowest order probe of gravitationally induced
non-Gaussianity. A useful version of it is the reduced bispectrum 
$Q$, which for the density field $\delta$ is given by  
\begin{equation}
Q_\delta%(\vec k_1, \vec k_2, \vec k_3) 
\equiv \frac{B_\delta(\vec k_1, \vec k_2, \vec k_3)}
{P_\delta(k_1)P_\delta(k_2) + P_\delta(k_2)P_\delta(k_3) + 
P_\delta(k_1)P_\delta(k_3) }. 
\end{equation}
$Q$ is useful because it is insensitive to the amplitude of the power
spectrum; thus e.g. at tree level and in the case of a scale free
linear power spectrum $P(k) \sim k^n$ it is static and scale
independent for regular gravity\cite{Bernardeau2002}.

\begin{figure}[tbp]
\renewcommand{\baselinestretch}{1}
\begin{center}
\begin{minipage}[b]{.45\textwidth}
\includegraphics[width=1\textwidth]{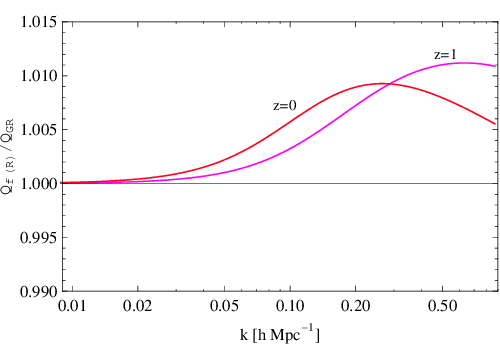}
\end{minipage}%

\vspace{-10pt}
\begin{minipage}[b]{.45\textwidth}
\includegraphics[width=1\textwidth]{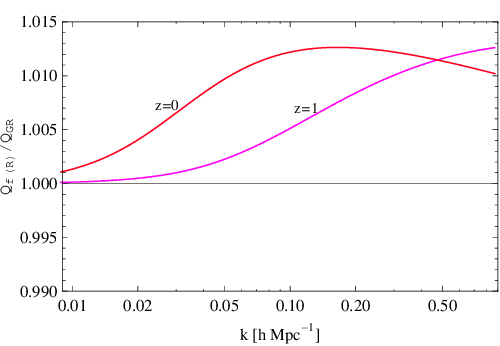}
\end{minipage}%

\vspace{-10pt}
\begin{minipage}[b]{.45\textwidth}
\includegraphics[width=1\textwidth]{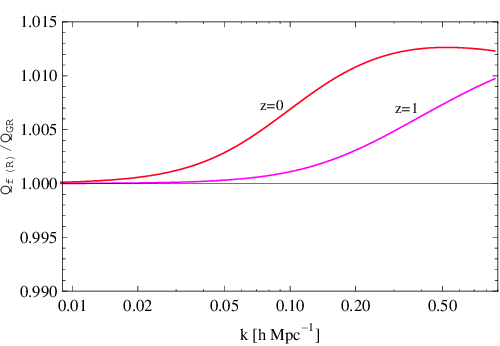}
\end{minipage}

\vspace{-10pt}
\end{center}

\caption{\label{fig:fig10}{\footnotesize The reduced bispectrum $Q$ for
    equilateral triangles for the $f(R)$ model
    compared to regular gravity. {\it Top panel:} $f_{R0}=10^{-5}$,
    $n=1$. {\it Middle panel:} $f_{R0}=10^{-4}$, $n=4$. {\it Bottom
    panel:}  $f_{R0}=10^{-5}$, $n=4$.}} 
\end{figure}

\section{Results}

In Fig. \ref{fig:fig3} we present the
 bispectrum for the $f(R)$ model 
 and its dependence on scale for two different redshifts. For comparison
we also show the bispectra predicted in GR. In the lower
 panel we present the ratio of the $f(R)$ bispectra to that in GR. 
%All calculations are performed at the
% lowest (tree) level of perturbation theory (that is we ignore all
% loop corrections). 
We have assumed the transfer function for the concordance
 $\Lambda-$CDM model consistent with the 5-year WMAP data \cite{2008arXiv0803.0586D}. Calculation is done at tree level with $\O_m = 0.24$.
%%BJ: Referenc e above
The bispectra in $f(R)$ gravity are enhanced relative to GR,
 increasingly so at high-$k$. The enhancement is of order 10-20\%
 for observationally relevant  scales around $k\sim 0.1$ and redshifts
 below unity.

We turn our attention to the reduced
bispectrum $Q$, which is expected to show features 
not associated with the linear power spectrum \cite{Bernardeau2002}. 
We show two relevant cases. The first one is with equilateral
triangles, shown in Fig. \ref{fig:fig10}. 
Deviations from GR are at the percent-level, which makes them impossible to
detect with current measurements. 
Qualitatively, the parameters of the model do influence the
scale and time dependence of the reduced bispectrum. 

\begin{figure}[tbp]
\renewcommand{\baselinestretch}{1}
\begin{center}
\leavevmode
%\epsfxsize=0.5\textwidth
%\epsffile{QFRRGOBTfR0=0.01n=1AdiabCDMv6a1.eps}
%\epsfxsize=0.5\textwidth
\includegraphics[width=0.45\textwidth]{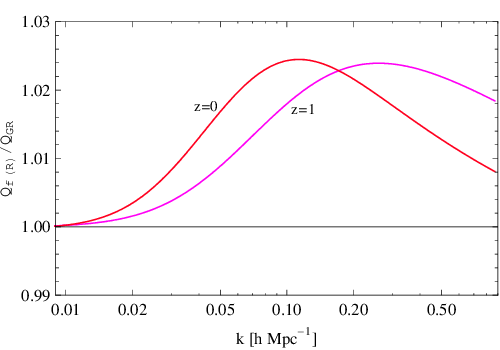}
\end{center}

\caption{\label{fig:fig12}{\footnotesize The reduced bispectrum $Q$ for
    isosceles triangles (ratio of sides lengths 1:3:3), 
    where the x-axis shows the length of the smallest $k$ in the triangle. 
    The ratio of $Q$ for the $f(R)$ model with $f_{R0}=10^{-5}$, $n=1$
    to regular  gravity is shown.}} 
\end{figure}

\begin{figure}[htbp]
\renewcommand{\baselinestretch}{1}
\begin{center}
\leavevmode
\begin{minipage}[b]{.45\textwidth}
\includegraphics[width=1\textwidth]{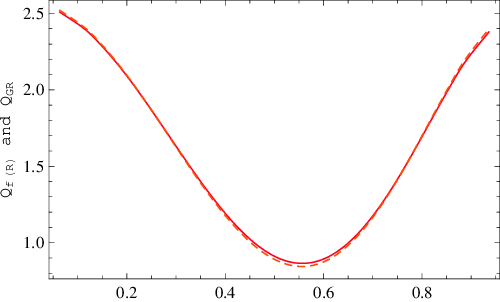}
\end{minipage}%

\vspace{-2pt}
\hspace{-7pt}
\begin{minipage}[b]{.457\textwidth}
\includegraphics[width=1\textwidth]{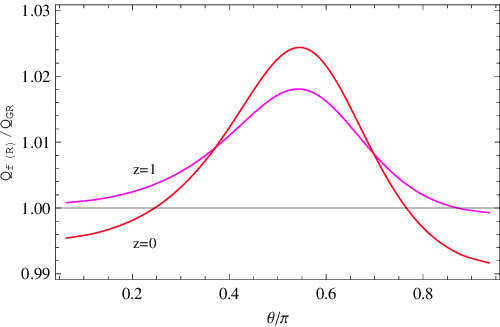}
\end{minipage}
\end{center}

\caption{\label{fig:fig11}{\footnotesize {\it Upper panel:} The
    angular dependence of the reduced bispectrum $Q$ for triangles
    with the two fixed side lengths in the ratio 1:3 and
    $k=0.1 h$/Mpc for the smaller one. 
The $f(R)$ model with $f_{R0}=10^{-5}$, $n=1$ at
    $z=0$ is used. The dashed line shows the prediction for GR. {\it Lower
    panel:} Comparison to regular gravity at two redshifts. We can
    observe that for obtuse shapes ($ \theta  \sim 0$) our model
    predicts a lower $Q$ for $f(R)$ gravity, while for shapes close to
    isosceles it is enhanced.}}  
\end{figure}

Slightly stronger deviations 
from regular gravity are observed for isosceles triangle configurations, shown
in Fig. \ref{fig:fig12}. Once again strong scale and time variation is
observed when changing the model parameters. We also 
show the angular dependence of the reduced bispectrum
for 3:1 ratio configurations and its comparison to regular gravity 
in Fig. \ref{fig:fig11}.

\subsection{Does the Linear Growth Factor Determine Nonlinear Clustering?}

Quasilinear effects, and the bispectrum in particular, show signatures
of gravitational clustering.  
However, we find that the reduced
bispectrum $Q$ remains very close to to that of $GR$ in the modified
gravity models we have considered (see also \cite{TatekawaTsujikawa} for a
related study).  This is qualitatively similar to previous
findings about the insensitivity of $Q$ to cosmological
parameters within a GR context \cite{Bernardeau2002}. 

Thus the deviations in the bispectrum are largely 
determined by the linear growth factor. 
Ambitious future surveys will be needed to achieve the percent level
measurements required to probe the unique signatures of 
modified gravity in the bispectrum. 
On the other hand, this means that the bispectrum can be used as a
consistency check on the power spectrum. It has been shown that the
bispectrum contains information comparable to the power spectrum, thus
improving the signal-to-noise \cite{2006PhRvD..74b3522S} . Equally
importantly, it is affected by sources of systematic error in
different ways and contains signatures of gravitational clustering
that are unlikely to be mimicked by other effects. 
%%BJ: Reference to Scoccimarro/Croce above

Currently the results of N-body simulations for
this f(R) model \cite{LimaHu} show that our assumptions are
consistent with their result for the power spectrum in the regime of
validity of perturbation theory (e.g. for $k \lesssim 0.3
[h/Mpc]$). On smaller scales, the chameleon mechanism modifies the
power spectrum. 

Modifications to the Newtonian potential 
were simulated by \cite{Stabenau2006} (also see \cite{Yukawa},\cite{Laszlo}).  
%%BJ: also cite Shirata et al and Liguori and Bean above - Shirata et al is included in the Yukawa citation, %AB. 
These simulation studies found that, to a good approximation, 
the nonlinear power spectra depend only on the initial conditions and
linear growth. The standard fitting formulae for Newtonian gravity
\cite{Smith2002} were adapted to predict the nonlinear
spectrum at a given redshift.  
Therefore tests for modified gravity using the power spectrum 
require either a measurement of
the scale dependent growth factor (in combination with the initial
spectrum measured from the CMB), or of measurements at multiple
epochs.  More likely a combination of probes will be used for robust
tests of gravity (see e.g. \cite{JZ}). 

Related studies of nonlinear clustering in $f(R)$ models or DGP
gravity are in progress \cite{Schmidt,Martino,Scoccimarro08}; these
authors are considering the power spectra, bispectra as well as halo
properties, in particular the halo mass function. The $f(R)$ studies of
\cite{LimaHu} and \cite{Schmidt} show distinct effects of the
chameleon field on small scales (comparable to galaxy and cluster
sized halos for most models); \cite{HuSaw2} suggest a fit for the nonlinear
power spectrum with additional parameters that describe the transition
to the small-scale regime. The DGP study of
\cite{Scoccimarro08} requires inclusion of nonlinear terms in the Poisson
equation. So clearly for different models there can be new
nonlinearities and couplings to additional fields that impact the small-scale
regime of clustering. Even so, for a class of models that includes the
$f(R)$ models studied here, the quasistatic, Newtonian description
parameterized by $\tilde{G}_{\rm eff}$ and $\eta=\Phi/\Psi$ applies
over a wide range of scales relevant for large-scale structure
observations. In this regime, to a good approximation,  many clustering
statistics can be predicted using the linear growth factor and the
standard gravity formalism. 

\subsection{Implications for Lensing and Dynamics}

%A very important issue is how some observational quantities are
%affected. 
\begin{figure}[htbp]
\renewcommand{\baselinestretch}{1}
\begin{center}
\leavevmode
%\epsfxsize=0.5\textwidth
%\epsffile{QFRRGOBTfR0=0.01n=1AdiabCDMv6a1.eps}
%\epsfxsize=0.5\textwidth
\includegraphics[width=0.45\textwidth]{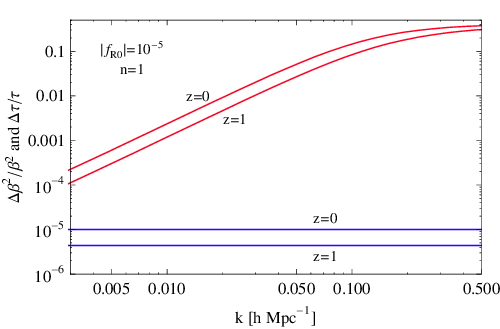}
\end{center}

\caption{\label{fig:fig13}{\footnotesize The fractional change in
    $\tau$ (blue lines) and $\beta^2$ (red lines) for the $f(R)$ model
    with $f_{R0}=10^{-5}$, $n=1$ compared to GR as a function of
    scale.}} 
\end{figure}

Lensing observations provide estimates of the  convergence power
spectrum and bispectrum (see e.g. \cite{Takada2004}). 
For a rough estimate of these quantities, we take
the source galaxies distribution to be a delta function at a given
redshift and take the lensing matter to be situated
at half the distance. Since we have taken the expansion
history to be $\L CDM$, so comoving
distances are the same as in GR. With these approximations, 
$P_{\k} \propto P_{\Phi - \Psi}$. From 
Eqn. \ref{eqn:MG3} we can see that the difference of the lensing
behavior of our modified gravity case and the regular gravity one is
given by (see also \cite{Bernardeau2004})
\begin{equation}
P_{\Phi - \Psi} \propto P_{\delta} \left( \frac{\tilde{G}_{\rm eff}(k,t_{1/2})}{G} \right)^2 
\end{equation}
where $t_{1/2}$ is the physical time at the redshift of the
lensing matter.  

Thus we can write: 
\begin{equation}
\frac{P_{\kappa MG}}{P_{\kappa GR}} \propto \frac{P_{\d MG}}{P_{\d GR}} \left( \frac{\tilde{G}_{\rm eff}(k,t_{1/2})}{G} \right)^2
\end{equation}

Analogously we can see that the ratio of convergence bispectra behaves
like $(\tilde{G}_{\rm eff}(k,t_{1/2})/G )^3$ 
but the ratio of the reduced bispectra behaves like
%$1/\tau$. 
\begin{equation}
 \frac{Q_{\kappa MG}}{Q_{\kappa GR}} \propto \frac{Q_{\d MG}}{Q_{\d
 GR}} \frac{G}{\tilde{G}_{\rm eff}(k,t_{1/2})}
%\frac{1}{\tau (k,a_{(z_{s}/2)})} 
\end{equation}

Thus the convergence power spectrum and reduced bispectrum could be
successfully used to differentiate and/or rule out models of modified
gravity. Unfortunately in the currently discussed model we have
Eqn. \ref{Gteff}: 
\begin{equation}
 \tau(k,t(a)) = \frac{\tilde{G}_{\rm eff}}{G} = \frac{1}{1+f_{R}(a)} 
\end{equation}
which deviates from unity at the order of $f_{R0}$, which is much
smaller than unity. Thus the convergence power spectrum and reduced
bispectrum follow almost identically the predictions for their matter
counterparts.  This means that the comparison of lensing to tracers of
mass fluctuations does not reveal distinct signatures of $f(R)$ gravity. 

The Newtonian potential
$\Psi$ drives dynamical observables such as 
%As explained in  Eqns. (27,28,29) it can be related
the redshift space power spectrum of galaxies \cite{JZ},
\cite{Zhang7}. The velocity growth factor
$D_{\theta}$ is  related to the density growth factor via the function
$\beta$: 
\begin{equation}
 D_{\theta} \propto a\beta H D
\end{equation}
where $\beta = d \ln D / d \ln a$.
This function varies with scale and expansion
factor for our $f(R)$ models. Both $\tau$ and $\beta$ can be seen in
Fig. \ref{fig:fig13} 
which clearly shows that observables based on peculiar velocities
would show a clear signature of $f(R)$ gravity. More detailed
calculations of lensing and velocity statistics are beyond the scope of
this study. 

\section{Discussion}

In this paper we studied the growth of structure in an $f(R)$ modified
gravity model \cite{HuSaw}.  
%which evades Solar system observations via its Chameleon mechanism. We used
In the quasilinear regime, we used perturbation theory to calculate the
three-point correlation function, the bispectrum, of matter. 
Our results are applicable up to scales at
which the derivation of the quasi-linear perturbation theory is still
valid. 

We found that in the bispectrum the dominant behavior is due to
the difference in the linear growth factors between the modified
gravity model and regular gravity. The bispectrum itself shows
significant departures in scale and time compared to the predictions
of GR. However the reduced bispectrum, which is
independent of the linear growth factor in perturbation theory for GR, 
remains within a few percent of the regular gravity prediction. 
It does show interesting signatures of modified gravity at the percent
level. 

Our results are consistent with studies of the nonlinear regime via
N-body simulations, which have found that on a wide range of 
scales the nonlinear power spectrum can be predicted using the
(modified gravity) linear growth factor in the 
standard formulae developed for Newtonian gravity. Our results imply
that three-point correlations follow this trend at the few percent
level. (The regime around and inside halos probed by \cite{LimaHu} 
to test the chameleon behavior is not included in our perturbative
study.) It would be interesting to compare perturbative and
simulation results for the bispectrum for the models considered here
and other modified gravity models. 

Upcoming surveys in the next five years will not attain the percent
level accuracy at which the reduced bispectrum shows distinct
signatures of $f(R)$ gravity. In this time-frame, the bispectrum will
be useful as a consistency check on potential deviations from GR found
in the power spectrum. Such a check is useful since 
measurement errors and scale dependent
biases of tracers can mimic some of the deviations in the power spectrum. 
Next generation surveys, to be carried out in the coming decade, will
provide sufficient accuracy to test the distinct signatures seen in
our results for the reduced bispectrum. With these surveys, the
bispectrum can provide truly new signatures of gravity. 

\bigskip

{\it Acknowledgments:}
We are grateful to Jacek Guzik, Wayne Hu, Mike Jarvis, Justin Khoury, 
Matt Martino, Fabian Schmidt, Roman Scoccimarro, Ravi Sheth, Fritz
Stabenau, Masahiro Takada and Pengjie Zhang.  BJ is supported in part
by NSF grant AST-0607667.

\end{document}